# Fake News Identification using Machine Learning Algorithms Based on Graph Features

Yuxuan Tian
Somerset, NJ
ytian27@seas.upenn.edu

*Abstract*—The spread of fake news has long been a social issue and the necessity of identifying it has become evident since its dangers are well recognized. In addition to causing uneasiness among the public, it has even more devastating consequences. For instance, it might lead to death during pandemics due to unverified medical instructions. This study aims to build a model for identifying fake news using graphs and machine learning algorithms. Instead of scanning the news content or user information, the research explicitly focuses on the spreading network, which shows the interconnection among people, and graph features such as the Eigenvector centrality, Jaccard Coefficient, and the shortest path. Fourteen features are extracted from graphs and tested in thirteen machine learning models. After analyzing these features and comparing the test result of machine learning models, the results reflect that propensity and centrality contribute highly to the classification. The best performing models reach 0.9913 and 0.9987 separately from datasets Twitter15 and Twitter16 using a modified tree classifier and Support Vector Classifier. This model can effectively predict fake news, prevent potential negative social impact caused by fake news, and provide a new perspective on graph feature selection for machine learning models.

I. Introduction

Composed of a large amount of user-generated content, the World Wide Web allows increasing information transportation on the Internet. With the trend of people sharing information, the amount of fake news is rising and spreading virally through websites, social media, and other informationsharing platforms. The credibility and reliability of the original source are not always verified, and therefore people have to analyze the information they receive in order to make informed responses. Fake news, or rumors, causes substantial negative social impact by confusing, threatening and misleading people. For example, during the COVID pandemic, scientists tried various methods to fight the infodemic of biased and fake medical news on the Internet to avoid unnecessary infections and the considerable cost in terms of lives lost and money spent (Cheng et al., 2021). The process of reading and spreading fake information occurs without people's awareness due to an inability to distinguish the accuracy of the content.

Many companies and websites are trying actively to reduce the spread of fake news. Some social media websites like Instagram will show alerts for all medical information during pandemics, helping to highlight material that needs to be carefully scrutinized. Websites like Snopes and Factcheck provide users with human-supervised labels on whether the news they are publishing is real. Researchers have developed many algorithms over the years. Most research done to classify fake news is based on the content itself. For example, Reis et al. (2019) used supervised machine learning with selected features like language, psycholinguistic and semantics to detect false information. Content-based features can be used for supervised machine learning. However, this approach is limited due to the lack of enough labeled samples, which are essential for supervised machine learning algorithms since a large amount of work is needed to process all contents.

II. Related Works

Researchers tried semi-supervised and unsupervised learning to test for fake news as well. Gangireddy et al. (2020) designed GTUT, a graph mining method using textual, user, and temporal information from network traces. They analyzed the bi-cliques network between users and the similarity between users and websites and expanded their research, achieving almost 80% accuracy. Benamira et al. (2020) proposed applying semi-supervised learning and graph neural networks to detect fake news, solely dependent on the news content. Therefore, to gain more information and to establish the overall structure besides the content, Lu and Li (2020) designed a Graphaware Co-Attention Networks (GCAN) model according to the time sequence of retweeting. Similarly, Nguyen et al. (2020) modeled Factual News Graph (FANG) to represent the social relationships among sources, news, and users and ran algorithms based on this structure, achieving better results than other methods in terms of accuracy.

Ma et al. (2017) published two datasets, Twitter15 and Twitter16, including real news and rumor information propagated, along with the routes and times they were propagated. They aimed to explore the patterns in the fake news transmitting and categorize news with four different tags: nonrumor, false rumor, true rumor, and unverified rumor. The use of a propagation tree kernel to predict the label of news and their results exceeds all former approaches.

Following their experiment, Bian et al. (2020) explored spreading and detecting rumors using Bi-Directional Graph Convolutional Networks. They constructed propagation and dispersion graphs and enhanced node features. By examining the graph from the original direction and opposite direction, they are able to get a better understanding of the overall structure.

Using clustering and cascading algorithms may help with the prediction of fake news by classification. Networks tend to form with homogeneous nodes or links, mainly caused by selective exposure to content. Vosoughi et al. (2018) state that fake news can diffuses deeper, faster, and farther than real news, providing the assumption for modeling cascading behaviors.

In this paper, I aim to predict the weight of edges in a weighted signed network using machine learning to identify fake information on the Internet. By adding nodes representing people, connecting edges representing spreading news, establishing features, and comparing machine learning algorithms performances, I hope to create an algorithm that can effectively detect the spread of fake news among Twitter users.

## III. BACKGROUND

To select features for the machine learning prediction models, I used various metrics which show the properties of the nodes and edges in a graph.

For each individual node, degree centrality can give a simple but effective measure of the centrality and can be calculated by counting the total number of connections linked to a node (Hansen et al., 2020). In addition, the Eigenvector centrality is added to the features. The Eigenvector centrality is based on the idea that people gain more centrality from some nodes over others (Golbeck, 2013). Mathematically, the eigenvector centrality of node $i$, the $i^{th}$ element of a vector $x$, can be defined as $Ax = \lambda x$ where $A$ is the adjacency matrix of graph $G$ with eigenvalue $\lambda$ (Phillip Bonacich, 1987).

I define the shortest path linking as following.

Function Shortest Path Linking ($Node_a$, $Node_b$):

Input: Two nodes, $Node_a$ and $Node_b$, where both of them are included in the graph
Output: The length of the shortest path between two nodes as an integer
IF Is not directed connected ($Node_a$, $Node_b$): RETURN Dijkstra shortest path ($G$, $Node_a$, $Node_b$) ELSE IF is directed connected ($Node_a$, $Node_b$):
Remove the edge between $Node_a$ and $Node_b$
RETURN Dijkstra shortest path ($G$, $Node_a$, $Node_b$)
Add an edge between $Node_a$ and $Node_b$

In addition to path lengths, I obtained the ratio of mutual sign neighbors using the algorithm presented below.

Function Mutual Sign Neighbors ($Node_a$, $Node_b$):

Input: Two nodes, $Node_a$ and $Node_b$, where both of them are included in the graph
Output: The ratio of the same sign nodes among the common neighbors of $Node_a$ and $Node_b$ as afloat.
K, Q = 0
FOR $Node_i$ IN all common neighbors ($Node_a$, $Node_b$): IF Edge weight ($Node_a$, $Node_i$) * Edge weight ($Node_b$, $Node_i$) ¿ 0:
K++
Q ++
RETURN K / Q

Moreover, I define the propensity of a node as the ratio of positive edges (real news) connected to a node:

Function Propensity ($Node_a$):

Input: One node, $Node_a$, where it is contained in the graph
Output: The fraction of edges that is positive as a float
K,Q = 0
FOR $Edge_i$ IN all _edges connected ($Node_a$):
IF Edge weight ($Edge_i$) >0:
K++
Q ++
RETURN K / Q

The Jaccard coefficient is a value between 0 and 1. It is calculated to represent the similarity between $u$ and $v$. It can be defined as the formula below, where $\tau(u)$ denotes the set of neighbors around $u$.

small small
= 
∩ 
$$J(u, v) = \frac{|\tau(u) \cap \tau(v)|}{|\tau(u) \cup \tau(v)|}$$

Fig. 1. Jaccard coefficient

Similarly, the Adamic Adar coefficient is used for measuring similarity between two nodes. However, instead of considering both union and intersection, it takes the inverse logarithmic degree centrality of the union of neighbors between $u$ and $v$ as the formula represented below (Liben-Nowell and Kleinberg, 2007). small small

$$A(u,v)= \sum_{w\in\tau(u)\cap\tau(v)} log|\tau(w)|$$

Fig. 2. Adamic Adar coefficient

Principal component analysis (PCA) and t-distributed stochastic neighbor embedding (tSNE) are two effective approaches to visualize and analyze the correlation among features. PCA creates new uncorrelated variables that could maximize variance, minimizing the information loss while increasing its interpretability (Jollife and Cadima, 2016). tSNE reaches the same goal but using manifold learning instead. It can process related high dimensional data spreading in multiple manifolds (van der Maaten and Hinton, 2008).

small

| Statistics of the dataset | Twitter15 | Twitter16 |
|---|---|---|
| Number of users | 276,663 | 173,487 |
| Number of source tweets | 1,490 | 818 |
| Number of false rumors | 370 | 205 |
| Number of true rumors | 372 | 205 |
| Average number of posts per user | 223 | 251 |

TABLE I STATISTICS OF DATASETS TWITTER15 AND TWITTER16

small

## IV. TECHNICAL APPROACH

Ma et al. (2017) provide datasets Twitter15 and Twitter16, which contain some news-related information from Twitter, including the news content, the label of the news (whether it is fake or real), users who retweet it and the order of retweeting. The datasets contain more than 2000 items of news for both fake and real news and more statistics information is presented in Table1. I build networks where users are nodes and users who tweet the same news will be connected by an edge. The edge attribute (the information that the edge contains) will belong to the news, the weight will be 1 for true information and -1 for fake information. NetworkX, Matplotlib and Gephi will be used to generate and visualize graphs (Hagberg et al., 2008; Hunter, 2007; Bastian et al., 2009).

To extract features for machine learning, I plan to run several graph algorithms exploring the attributes of both nodes and edges. From the level of nodes, I plan to use in-degrees, degree centrality, eigenvector centrality, clustering coefficient and propensity as features of the machine learning model. For edges, I will calculate the shortest path between two endpoints and the Jaccard coefficient and the Adamic Adar coefficient for this pair of endpoints.

Nodes on both sides of the edges can offer extra information to the edge since I am simulating the interaction of news spreading between two people using edges. Different features, including clustering coefficients and centrality coefficients of nodes, are calculated. The ratio of neighbors who have the same sign to all neighbors is essential to the classification due to the fact that people have a higher possibility of spreading fake news if their friends or connected people are spreading fake news. The ratio can be calculated by getting the weights of the edges to which the two nodes are connected. The Jaccard coefficient and the Adamic Adar coefficient indicate the similarity of the two nodes.

After finishing features selection, I then compare and contrast machine learning classification algorithms to the final model using Scikit-learn, which offers accessible models (Pedregosa et al., 2011). Naïve Bayes can be used to build, run, and analyze both the Gaussian and the Kernel Bayes' classifiers. Decision trees can be effective while explainable, and therefore I try different splits, including 100 splits, 25 splits and 4 splits, height, and regularization methods. I will experimentally analyze their performance. Tree ensemble methods might be implemented if needed, including boosted tree, bagged tree, and RUSBoosted tree. This choice is because Dietterich (2000) demonstrates that ensemble methods using sets of classifiers often have better performance than single classifier methods.

Different combinations of features and algorithms will be tested and analyzed for predicting the fake news on the Twitter15 dataset.

## V. EXPERIMENT

The experiment is composed of graph modeling, feature selection and analysis, and machine learning model training.

### A. Graph Modeling

Utilizing the regular expression and string processing strategies, I extracted the id and label of news in the Twitter15 dataset and the list of users who retweet them from 1491 txt files (1 for the label and 1490 for each news source). Then, I created a signed undirected graph based on the dataset, which I assigned a positive integer one for authentic news and a negative one for rumors. After dropping the self-loop edges and completely isolating nodes, the graph's statistics are obtained in Table 2. Then, the same processes are applied to Twitter16 as well.

|  | Twitter15 | Twitter16 |
|---|---|---|
| Number of nodes | 223346 | 144324 |
| Number of edges | 250126 | 158842 |
| Average degree | 2.2398 | 2.2012 |

TABLE II
STATISTICS OF NETWORKS

### B. Feature Selections

I am able to obtain the information from both nodes and edges. The target value that I aim to predict is the weight of edges. The weight of 1 means the news corresponds to real news, while -1 stands for fake news. All features in the machine learning model should contribute to predicting the weight of edges, which indicates whether the news is fake. Therefore, I select 14 features in total, as shown in Table 3.

| In-degree of Node a | In-degree of Node b |
|---|---|
| Node a centrality | Node b centrality |
| Node a eigenvector centrality | Node b eigenvector centrality |
| Node a clustering coefficient | Node b clustering coefficient |
| Node a propensity | Node b propensity |
| Fraction of mutual signed neighbors | Shortest path |
| Jaccard coefficient | Adamic Adar coefficient |

TABLE III
FEATURES EXTRACTED

Five pairs of the features belong to two nodes connected to an edge. From the degree of the node and degree centrality index, I can evaluate the node's role in the network, which might affect the reliability of the spreading news. Eigenvector centrality is also used since the importance of nodes depends not only on the number of neighbors but also on the importance of neighbors of that node. In addition, the clustering coefficient is used to describe the ability of the node to group with others. It is helpful for the prediction since people who tightly connect to a community sometimes have quite different personalities than those who connect with friends and communities loosely or not at all. Personality impacts decisions such as whether to believe and spread rumors largely when the information is unreliable (Carter et al., 2013). More importantly, since I am predicting the weight of edges, I define propensity as the fraction of positive edges among all edges the node is connected to. For example, a person who connects mostly with fake news has a greater possibility of spreading fake news than someone who only has access to real news.

Aside from pairwise node features, I collect four other kinds of information between nodes. First, I calculate the shortest path except the directed connected path using Dijkstra's algorithm. The shortest path between two nodes can indicate the size of the impact of the creation of a new edge. For example, two nodes that are not connected or very far away from each other shows that the current edge is relatively essential to communication and might be the bridge between two clusters. Unlike other nodes inside clusters, these essential nodes might have a higher or lower possibility of spreading fake news. Moreover, the fraction of mutual same sign neighbors is included in one of the features. As I defined in the background section, this feature is extracted by finding the common neighbors of both nodes and calculating the fraction of common signs they have with them. The consistencies of two nodes might reveal the similarity of two nodes and further contribute to the prediction task. The Jaccard coefficient and the Adarmic Adar coefficient are also collected to describe the similarities between two nodes, and these two can present information from some additional perspectives.

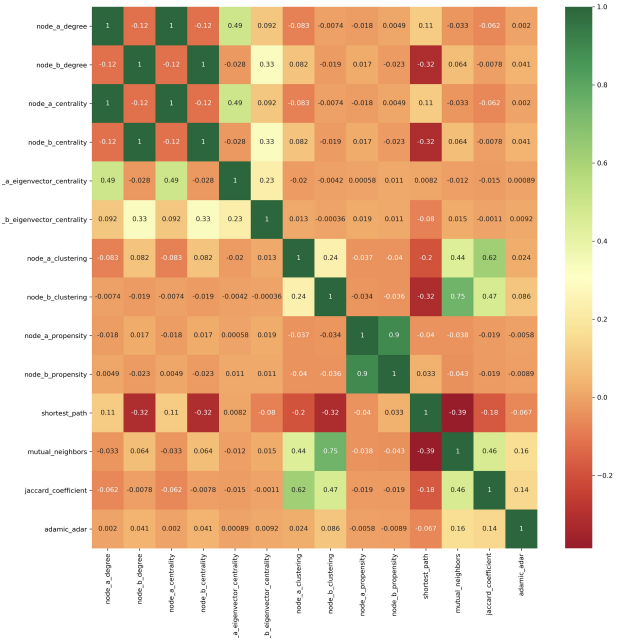

Fig. 3. Feature correlation for the Twitter15 dataset

Principal Component Analysis (PCA) and t-distributed stochastic neighbor embedding (tSNE) are used to better visualize the dataset by dimensionality reduction based on the correlation analysis (van der Maaten and Hinton, 2008). Graphs shown in Figure 3 and Figure 4 are the correlation plots for the features in both the Twitter15 and Twitter16 datasets. Since the nodes' degree centrality is directed propositional to the degree of the node, I can notice that the top-left corner has a 4x4 matrix that contains several 1s, which indicate directed correlation. Similarly, eigenvector centrality also considers degree, which explains why the value between eigenvector centrality and degrees is high. Interestingly, the correlation between the propensity of the two nodes is 0.86 and 0.9 for Twitter15 and Twitter16, respectively, which reveals the idea that people tend to communicate with people who have a similar propensity. The correlation between the ratio of mutual sign neighbors and the clustering coefficient of the node is high as well. That can explain why a person with a higher clustering coefficient tends to belong to a group and therefore have a higher ratio of mutual sign neighbors.

Since these features are interconnected, I ran the PCA algorithm. PCA can reduce the dimension of the datasets by transforming it into a new set of variables called uncorrelated principal components. Using only two principal components, I can obtain more than 70% of the original information based on the generated screen plots (Figure 5). Therefore, I plot the fake news with weight -1 as blue and real news with weight 1 as yellow on a two-dimensional space. It is not hard to divide the sample points into two parts visually though it has some intersections as shown in Figure 6.

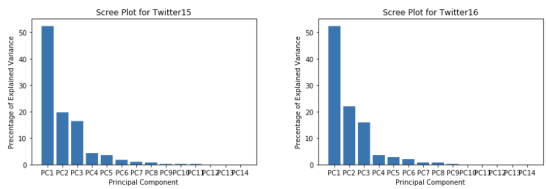

Fig. 4. Scree Plot for PCA analysis for datasets

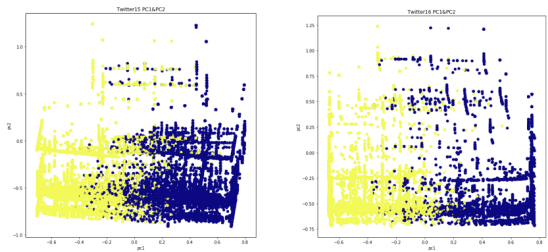

Fig. 5. Graph based on first and second principal component

Different from PCA, tSNE uses manifold learning, a topological space that locally resembles Euclid's space near each point, to reduce the dimensions. More specifically, it converts points similarity to joint probability and aims to reduce divergence (Maaten and Hinton, 2008). Also, since these datasets only contain 14 features, I do not need to be concerned about the noises brought by adding extra dimensions. I drew tSNE graphs separately using 5000, 10000 and 50000 samples, and some patterns are shown in Figure 6. With the increment of sample size, the spreading patterns of fake and real news are switching in the two-dimensional space. Though the distance between data points in the two-dimensional space is not necessarily able to be interpreted, the distribution and patterns might indicate that the data can be separated through terpreted, the distribution and patterns might indicate that the data can be separated through some techniques back in high-dimensional space.

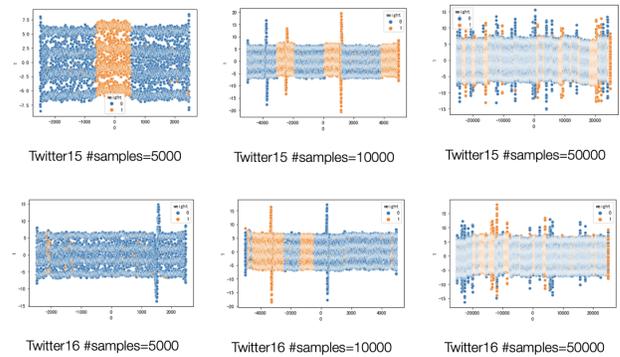

Fig. 6. tSNE visualization for both datasets in different sample sizes

### C. Machine Learning Preprocessing and Models

After collecting, preparing and analyzing features, I tested the datasets in several linear models, tree models, Support Vector Machine and other models. Datasets are normalized using MinMaxScaler using the formula below. Since some features like node propensities and clustering coefficients are scaled in zero to one beforehand, and some other features like degrees and shortest path have not been scaled yet. The MinMaxScaler normalization might lead to fast convergence and a good chance of getting the optimal solution. small small Since the labels are fairly well balanced, with

$$X_{scaled} = \frac{X - X_{min}}{X_{max} - X_{min}}$$

Fig. 7. MinMaxScaler

about 50% of real news and the rest fake, accuracy is used to compare all machine learning algorithms. Accuracy here is explicitly defined as the number of correction predictions over a total number of predictions. Or accuracy can be defined as following using the truth table, where TP, TN, FP, and FN stand for true positive, true negative, false positive and false negative, respectively.

small small Logistic regression is based on the idea of

$$Accuracy = \frac{TP+TN}{TP+TN+FP+FN}$$

Fig. 8. Accuracy

finding the best fit subspace to predict the probability (between 0 and 1). The Sigmoid function (Figure 10) is used to map the result and finish the classification task.

small small The accuracy of the logistic regression algorithm is 0.9893 and 0.9948 for the two datasets, Twitter15 and Twitter16, separately. Assuming the data follows Gaussian distribution and Bernoulli distribution, I tested the Naive Bayes algorithm with both assumptions and the accuracy for
Twitter15 is 0.9846 and 0.9819 and Twitter 16 is 0.9897 and 1

$$g(z) = \frac{1}{1+e^{(-z)}}$$

Fig. 9. Sigmoid function

0.9902. In comparison, logistic regression has slightly better performance over Naive Bayes. This can be explained by many features in the dataset and is not independent as shown in PCA analysis. Naive Bayes has overly confident assumptions on the independence of the features and processes them separately without thoroughly analyzing the overlapping information.

Ensemble models are used, specifically AdaBoost, an algorithm that fits a sequence of weak learners, or machine learning models with high uncertainty, and repeatedly votes to produce the final result (Freund and Schapire, 1997). Several weak classifiers might learn from other's wrongly classified samples and hence could establish a strong ensemble model. After adjusting parameters and testing results, the best-integrated model reaches 0.9912 and 0.9963 accuracy individually.

To better explain, decision tree classifiers are used with different levels of maximum splits (4, 25, 100, and infinite). The accuracies for dataset Twitter15 are 0.9878, 0.9910, 0.9913 and 0.9908 and for dataset Twitter16 are 0.9948, 0.9960, 0.9959 and 0.9958 for four different maximum splits. As Figure 11 indicates, node propensity and eigenvector centrality play the two most essential roles in prediction tree models. People who are exposed to fake news in their communities are more likely to spread it to others.

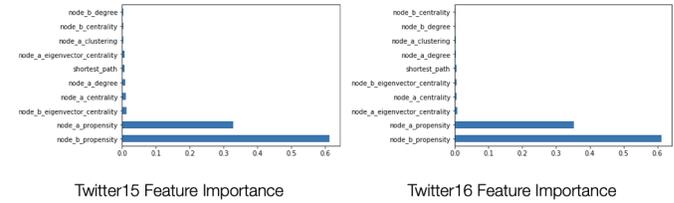

Fig. 10. Ten most important features in tree algorithms

Lastly, several Support Vector Machine models with different kernels are utilized since the datasets seem not linearly separable according to the result of PCA. Therefore, using SVM kernels might convert the datasets to higher dimensions and make them separable. I experimented with six kernel models, including the linear kernel, Linear SVC, RBF kernel, polynomial (degree 3) kernel, SVC with polynomial (degree 5) kernel, and SVC with polynomial (degree 7) kernel. The accuracies for Twitter 15 are 0.9892, 0.9889, 0.9900, 0.9893, 0.9867 and 0.9811 and Twitter16 are 0.9977, 0.9960, 0.9983,
0.9977, 0.9987 and 0.9967.

VI. CONCLUSIONS

My work collects and analyzes the information on the news spreading networks. Different features and machine learning models are tested, and the best average accuracies among these algorithms reach 0.9913 and 0.9987 for datasets Twitter15 and Twitter16 using modified decision tree algorithms and the Support Vector Machine. From feature analysis, I discover that people with similar propensity tend to create connections, and the clustering coefficient is proportional to mutually signed neighbors in these datasets. The complete results are summarized in Table 5. This finding can be applied to spread the information and for public communication. Moreover, the experiment shows the importance of propensity and centrality, and other models can adopt it. Future studies should test it on more datasets with some additional machine learning models, including neural networks and graph algorithms such as PageRank, to obtain more generalized results. Based on these conclusions, social media platforms should consider implementing network-level algorithms to detect fake news if necessary to prevent potential negative social impact.

| Method/Accuracy | Twitter15 | Twitter16 |
| --- | --- | --- |
| Logistic Regression | 0.9893 | 0.9948 |
| Naive Bayes | 0.9846 | 0.9819 |
| AdaBoost | 0.9912 | 0.9963 |
| Coarse Tree(4 max splits) | 0.9878 | 0.9948 |
| Median Tree(25 max splits) | 0.9910 | 0.9960 |
| Fine Tree(100 max splits) | 0.9913 | 0.9959 |
| Very Fine Tree(no restricted splits) | 0.9908 | 0.9958 |
| SVM linear kernel | 0.9892 | 0.9977 |
| Linear SVC | 0.9889 | 0.9960 |
| SVM RBF kernel | 0.9900 | 0.9983 |
| SVC with polynomial (degree 3) kernel | 0.9893 | 0.9977 |
| SVC with polynomial (degree 5) kernel | 0.9867 | 0.9987 |
| SVC with polynomial (degree 7) kernel | 0.9811 | 0.9967 |

TABLE IV
ACCURACY SUMMARY FOR ALL ALGORITHMS TESTED